\documentclass[epsf,twocolumn,showpacs,preprintnumbers]{revtex4}
\usepackage{graphics}
\usepackage{graphicx}
\usepackage{dcolumn} 
\usepackage{bm}
\usepackage{color}
\usepackage{epsfig}
\usepackage{comment}
\newcommand{\bfpo}{BaFe$_2$(PO$_4$)$_2$}
\begin{document}
\title{Design of Chern Insulating Phases in  Honeycomb Lattices
}
\author{Warren E. Pickett$^1$}
\email{pickett@physics.ucdavis.edu}
\author{Kwan-Woo Lee$^{2,3}$}
\email{mckwan@korea.ac.kr}
\author{Rossitza Pentcheva$^4$}
\email{rossitza.pentcheva@uni-due.de}
\affiliation{
 $^1$Department of Physics, University of California, Davis,
  CA 95616, USA
 $^2$Department of Applied Physics, Graduate School, Korea University, Sejong 30019, Korea\\
 $^3$ Department of Display and Semiconductor Physics, Korea University, Sejong 30019, Korea \\
 $^4$Department of Physics and Center for Nanointegration Duisburg-Essen (CENIDE),
   University of Duisburg-Essen, 47057 Duisburg, Germany,\\
}
\date{\today}

\begin{abstract}
The search for robust examples of the magnetic version of topological insulators,
referred to as quantum anomalous Hall insulators or simply Chern insulators,
so far lacks success. Our groups have explored two distinct possibilities
based on multiorbital $3d$
oxide honeycomb lattices. Each has a
Chern insulating phase near the ground state, but materials parameters were
not appropriate to produce a viable Chern insulator. Further exploration of
 one of these classes, by substituting open shell $3d$ with $4d$ and $5d$ counterparts, 
has led to realistic prediction of Chern insulating ground states. Here
we recount the design process, discussing the many energy scales that are
active in participating (or resisting) the desired Chern insulator phase. 
\end{abstract}
\maketitle

\section{Introduction}
In recent years the search for new materials with properties or electronic phases
of interest has experienced a change of paradigm. The conventional
method, referred to in the U.S. as the Edisonian method, is by (informed) trial and error.
In fact Thomas Edison went as far as possible beyond the simplistic ``trial
and error" approach that is often brought to mind. He searched carefully in
the databases that existed, he gave close attention to learning from earlier
(successful, but also those dubbed failed) experiments, and he worked
unceasingly toward specific goals -- in his case, marketable properties and
products. His approach was far from trial and error. 

The recent paradigm change in the U.S. is promoted by the Materials Genome
Initiative, a government-wide program to introduce computational
expertise intimately into the materials development effort. The essence is to integrate
computational capabilities into an integrated loop: computational design,
synthesis of predicted materials, and characterization of properties to test whether
desired aspects are present, with each point of the loop feeding back and forth with the
computational node. Emphasis on computation is based on (i) the persistent and increasingly
impressive progression of computer power, (ii) the demonstration that 
materials theory and its computational implementation leads to reliable
predictions, and (iii) that this information (materials data) can be produced at a fraction of
the cost and effort of laboratory exploration. 
Related efforts at computational design are active in other countries, judging
from published papers.

\begin{figure}[tbp]
{\resizebox{9cm}{8cm}{\includegraphics{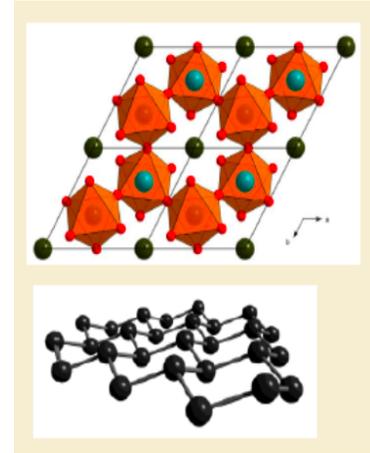}}}
\vskip -10mm
\caption{(Color online) (top) View from above of the bilayer of (111) LaXO$_3$,
illustrating the corner-sharing XO$_6$ octahedra. Alternate octahedra
are centered at different heights. (bottom) Perspective view of the
buckled honeycomb X bilayer sublattice.
}
\label{str}
\end{figure}

One outgrowth of this new paradigm is the ``high throughput'' approach:\cite{Ceder} choose a
viable computational algorithm, identify a {\it descriptor} (or a few), that is, a computable
property that is essential to the desired application, then to search classes of materials
thought to be likely candidates -- this may be thousands of materials --
and sort through results to find promising candidates. This approach 
can be successful when the number of degrees of freedom in the materials is 
understood and limited,
when the descriptor is clear, and the algorithm can be automated. Access to
considerable computer time can often be obtained, and the procedure can be
implemented and optimized.

Some important problems cannot be approached so directly. Suppose one wishes
to find, or design, a new high temperature superconductor, following the 
leads of the cuprates and Fe-pnictides and -chalcogenides which neighbor
magnetic phases. This is really challenging, since there is no theory in
these two classes of superconductors that specifies just what descriptor(s) one
should adopt. In these attempts, choices focus on similarities of crystal 
structure or electronic structure, or on
similar magnetic phases neighboring the superconductors. The few efforts in
this area have not yet produced positive results, though they are
still in the early stage.  

In the last few decades, the two-dimensional (2D) honeycomb lattice has attracted 
interest due to several exotic physical phenomena related with 
superconductivity, novel magnetic states,
and topological phases, some of them building on the Dirac point 2D material graphene.  
Recently, discoveries of topological insulators, originally on the honeycomb
lattice,\cite{kane} have stimulated 
increased research on properties of systems with honeycomb 
lattices.\cite{fiete,shitade,zhou,nandk}

The design and discovery of Chern insulators, the magnetic version of topological
insulators (TIs), is one of the current forefront pursuits in research on topological
materials. Unlike their time-reversal symmetric cousins, the non-magnetic
$Z_2$ TIs characterized\cite{kane} by topological indices $\nu_0(\nu_1 \nu_2 \nu_3)$ which
can be only zero or one, the Chern number ${\cal C}$ can be any integer. Aside from
the obvious interest, and complication, of being a magnetic system, this unbounded
aspect of values of ${\cal C}$ provides new topological behavior to explore and
perhaps exploit. One clear feature is that, while the bulk is insulating, there
are ${\cal C}$ boundary bands crossing the gap. These bands provide the quantum
anomalous Hall effect, thus a larger ${\cal C}$ will provide a correspondingly
larger Hall response.   

The initial expectation of realizing a Chern insulator was to incorporate some
magnetic ions onto or within a TI,\cite{earlychern} or layering transition
metal oxides.\cite{Garrity,Baidya2016,Wang2017} 
Several hurdles have 
impeded the first approach.
First, most known TIs have very small gaps (a few meV), complicating the studies.
Second, it is unfortunate that known TIs are not truly insulating in the bulk
 due to defects, hence the TI-specific boundary state behavior becomes challenging
to separate from bulk contributions. Third, several proposals have suggested 
submonolayers of magnetic ions or nonstoichiometric additions of magnetic ions,
whereby disorder and non-stoichiometry provide experimental challenges.
Synthesis and characterization of such materials presents its own challenges.

These complications and limitations may be alleviated by turning to (1)
stoichiometric systems, and (2) wider gap materials. For these reasons,
transition metal oxides (TMOs) provide promise. Bandgaps can be expected to be larger,
and a great deal of expertise in the synthesis of both bulk and layer-by-layer grown
oxides has accumulated.  (Even so, Chern insulators built on TMOs also may have only
small gaps.\cite{pardo2009,huang2015,cai2015})  
Along with these positive aspects, TMOs provide a
grand palette of possibilities, many of which offer interplay 
among lattice, spin, and orbital degrees of freedom, together with correlation effects 
and spin-orbit coupling (SOC), thus providing an enormous field to explore and exploit.
Along with this large playing field comes a variety of challenges. Here we discuss
succinctly our experience in the context of two honeycomb lattice systems.

\section{Two Honeycomb Lattices}
The two lattices we have studied display some differences. First, we
should mention commonalities. The ideal honeycomb
lattice has features that distinguish it from the common square lattice. It
has two sites per cell (sublattices A and B), making it intrinsically
multiorbital. The sublattice (``valley'') degree of freedom has been exploited
widely in the study of graphene. The local symmetry of each site is threefold,
which incorporates its own aspects. One is that the $t_{2g}$ subshell on a
transition metal ion is represented conveniently by the linear combinations
\begin{eqnarray}
\phi_m=\frac{1}{\sqrt{3}}(\xi_m^0d_{xy}+\xi_m^1d_{yz}+\xi_m^2d_{zx}),
\label{eq1}
\end{eqnarray}
where the phase factor is $\xi_m=exp(i\frac{2\pi m}{3})$, $m=<L_z>$
is the projection of the orbital moment, and the superscript is
an exponent.
There is the fully symmetric (with respect to threefold
rotations) member $a_{1g}$ ($m$=0) and a pair of 
complex $e_g^{\prime}$ orbitals ($m=\pm 1$).
This distinction from real $xy,yz,zx$ orbitals becomes 
crucial\cite{fiete} when SOC 
is active, and more especially so for the magnetic ions we are pursuing.

Both systems we have studied have open shell transition metal ions on the
2D honeycomb lattice sites, so each sublattice itself is multiorbital.
Experience in related TMOs has been important.\cite{bacro3,doennig2,doennig}
We note that the density functional theory for correlated materials DFT+$U$
method is necessary to model the gap in transition metal oxides, here $U$ is the
intra-atomic Coulomb repulsion (Hubbard) energy. We have used all-electron, 
full potential methods in our studies.\cite{fplo1,wien2k}~
\bfpo~ has the ideal honeycomb Fe sublattice\cite{bfpo1,bfpo2,bfpo3} 
for which sublattice symmetry
breaking is not an important factor.\cite{BFPO1,BFPO2} 
The other is comprised of a (111)
bilayer of $X$ cations encased in the perovskite 
insulator LaAlO$_3$ (denoted 2LXO/LAO), 
forming a buckled honeycomb lattice,\cite{xiao2011,cook} illustrated in Fig. 1.
The transition metal $X$ ion
sits, in the idealized structure, at a site of local cubic symmetry
within an O$_6$ octahedron, but further neighbors (or structural relaxation)
reduces symmetry to trigonal and even lower, as broken symmetries 
proliferate. 
Transition metal ions are prone to develop magnetic moments, after which
SOC will induce an orbital magnetic moment.\cite{2LXOb,2LXOa}
Breaking of spin, orbital, sublattice, etc. symmetry is accompanied
by geometric distortion and perhaps caused by it, enlarging the set of
degrees of freedom. These degrees include charge order, spin imbalance, 
orbital polarization
and orbital ordering, structural distortion, and perhaps more exotic quantum
order parameters although we have not pursued that possibility.

The energy scales are several, providing rich behavior but challenges for
material design. The bandwidth $W$ of the active orbitals;
on-site Coulomb repulsion $U$; Hund's magnetic coupling $J$; crystal
field splitting $\Delta_{cf}$; trigonal crystal subfield splitting
$\delta_{cf}$; SOC coupling strength $\xi$; Jahn-Teller distortion
energy(s). Evidently the complications
are many. In such a broad field of possibilities, it is beyond current
understanding to define a descriptor, or a few of them, that will allow a
computer code to evaluate whether a given case is near a Chern insulator,
or not. This is a problem that invites intimate human-machine interaction.

\subsection{Chern-ness in the Ising Ferromagnet BFPO}
To summarize briefly: in \bfpo ~it was necessary to include $U$ together with SOC
to realize any appreciable gap. A transition from
Chern insulator at small $U$ to trivial Mott insulator at larger $U$ 
occurred at a bandgap closing and reopening for the critical value
$U_c$=2.45 eV. Since calculated values of $U$ for Fe are 4-5 eV, our
conclusion was that the ground state is a conventional Mott insulator. It is
noteworthy that the 
topological transition coincides\cite{BFPO1,BFPO2} with a transition from a small orbital
moment on Fe to a value of 0.7-0.8 $\mu_B$, an extremely large value for
a $3d$ ion and near the 1 $\mu_B$ limit for the $t_{2g}$ subshell. It
is natural to suspect a connection, if not a causal relationship. 
This large moment
is accompanied by a very large calculated magnetocrystalline anisotropy
energy, which also accounts for the observed Ising nature of the ferromagnetic
state.

\subsection{Chern-ness in the 2LXO/LAO system}
Our initial focus on $3d$ ions led  us to obtain the ground states
for the entire $3d$ series 2LXO/LAO, $X$=Ti, V, Cr, Mn, Fe, Co, Ni, Cu. All are
magnetic, and those with filled subshells are structurally stable and
uninteresting. An open subshell led, depending on band filling, to
charge ordering, orbital ordering, and in some cases structural distortion,
all typical Mott behavior.
Before accounting for SOC, the Ti, Mn, and Co cases displayed Dirac
points in their majority band structure. SOC broke
symmetry and resulted in gap opening, closely associated with the
breaking of sublattice symmetry by the various order parameters.
These systems relaxed through a Chern insulating phase, but the
distortions were too large; before becoming fully relaxed, the
gap closed and band inversion disappeared, and the ground states 
were conventional, albeit symmetry broken Mott insulators. 

The reasoning -- the resulting design principles -- will be
discussed a little more in the next section. An evident conclusion
was that the distortion, though unavoidable in this system, is
not destructive but is too severe. 
This tendency could be combatted by decreasing the
tendency toward Jahn-Teller distortion, which is related to
localization of the orbitals, and by increasing SOC strength
which would serve to retain band entanglement to larger gap
values. Calculations extended to the $4d$ and $5d$ cations
that followed led to the prediction\cite{2LXOa}
of Chern insulating ground states in the $4d$ $X$=Ru compound
(Chern number ${\cal C}$=-1, a gap of 130 meV) and the $5d$ 
Chern insulator $X$=Os (${\cal C}$=2, gap of 50 meV). It is
noteworthy that each of these has a modest value of orbital 
moment of 0.16-0.19 $\mu_B$, especially since magnetic $5d$ 
ions have shown many times to display larger orbital moments.

\section{Discussion and Summary}
Chern insulators arise from (ferro)magnetic materials in which spin-orbit
interactions open a gap, leaving ``inverted bands.'' This last item is
not transparent in the electronic structure; one simply must calculate the
Berry curvature and integrate it over the zone, and if it is non-zero 
(it will be an integer) then the bands are inverted. The character of the
bands that is essential in providing a fruitful Berry connection needs
further study. When it is a sum over occupied bands (which is common)
the critical characteristic -- a topological one -- is even more elusive.
 
Our analysis of the evolution of the electronic system, versus interaction
strength $U$ for \bfpo, and versus structural distortion (related to $U$)
provided the necessary
guidance for successful design of Chern phases in buckled honeycomb lattices, 
but unfortunately without making the microscopic mechanism evident. The
emerging design principles became for us\\
$\bullet$ study honeycomb lattices. This is not a criterion, but they 
provide an intrinsic richness not shared by simpler, single site lattices.\\
$\bullet$ focus on systems that promote a Dirac
point at $K$ in the symmetric phase; recall that a Dirac point provides
a singularity in the topological nature of the system, \\
$\bullet$ avoid large symmetry-breaking effects such as the
Jahn-Teller distortions that open large gaps and destroyed the Chern
phases in the $3d$ systems, \\
$\bullet$ vary the strain to obtain optimal values, because electronic configurations are
sensitive to strain,  \\
$\bullet$ increase the SOC strength, as it sustains the
band entanglement over a proportionately greater energy and can be expected
to retain entanglement with larger gap.  

The Dirac point aspect seems to be one unifying underlying feature
in the systems we have studied. 
The interaction strength $U$, which
also tends to enforce anisotropic, Jahn-Teller-active ions, appears to be
a strong player.  Sensitivity to strain: this item reflects the many
studies which reveal that strain can strongly influence electronic systems,
by providing symmetry-breaking forces that spur, for example, orbital
polarization with its important consequences.
An increase in SOC strength has been widely postulated as a positive
influence, making $5d$ systems one focus of searches for Chern insulators.
These concepts are guiding our further studies on this topic.

{\it Acknowledgments.}`
 This research was supported by National Research Foundation of Korea 
 Grant No. NRF-2016R1A2B4009579 (K.W.L),
 by U.S. NSF DMREF Grant DMR-1534719 (W.E.P.), and by
 the German Science Foundation within SFB/TR80, project G3 (R.P.).


\begin{thebibliography}{10}

\bibitem{Ceder}	
D. Morgan, G. Ceder, S. Curtarolo,
High-throughput and data mining with ab initio methods,
Measurement Sci. and Tech. {\bf 16}, 296 (2005).

\bibitem{kane} C. L. Kane and E. J. Mele,
 $Z_2$ Topological Order and the Quantum Spin Hall Effect.
 Phys. Rev. Lett. {\bf 95}, 146802 (2005).


\bibitem{fiete} A. R\"{u}egg and G. A. Fiete,
 Topological insulators from complex orbital order
 in transition-metal oxides heterostructures.
 Phys. Rev. B {\bf 84}, 201103(R) (2011).

\bibitem{shitade} A. Shitade, H. Katsura, J. Kune\v{s}, X.-L. Qi,
 S.-C. Zhang, and N. Nagaosa,
 Quantum Spin Hall Effect in a Transition Metal Oxide Na$_2$IrO$_3$.
 Phys. Rev. Lett. {\bf 102}, 256403 (2009).

\bibitem{zhou} M. Zhou, Z. Liu, W. Ming, Z. Wang, and F. Liu,
 $sd^2$ Graphene: Kagome Band in a Hexagonal Lattice,
 Phys. Rev. Lett. {\bf 113}, 236802 (2014).

\bibitem{nandk} R. Nandkishore, L. S. Levitov, and A. V. Chubukov,
 Chiral superconductivity from repulsive interactions in doped graphene,
  Nature Phys. {\bf 8}, 158 (2012).

\bibitem{earlychern}C.-X. Liu, X.-L. Qi, X. Dai, Z. Fang, and S.-C. Zhang,
Phys. Rev. Lett. {\bf 101}, 146802 (2008).

\bibitem{Garrity}K. F. Garrity and D. Vanderbilt,
Chern insulators from heavy atoms on magnetic substrates,
Phys. Rev. Lett. {\bf 110}, 116802 (2013).

\bibitem{Baidya2016}S. Baidya, U. V. Waghmere, A. Paramekanti, and
T. Saha-Dasgupta, High-temperature large-gap quantum anomalous Hall insulating
state in ultrathin double perovskite films, Phys. Rev. B {\bf 94},
  155405 (2016).
\bibitem{Wang2017}H. P. Wang, W. Luo, and H. J. Xiang,
 Prediction of high-temperature quantum anomalous Hall effect in
 two-dimensional transition metal oxides,
 Phys. Rev. B {\bf 95}, 125430 (2017).

\bibitem{pardo2009} V. Pardo and W. E. Pickett,
  Half-metallic semi-Dirac Point Generated by Quantum
 Confinement in TiO$_2$/VO$_2$ Nanostructures,
 Phys. Rev. Lett. {\bf 102}, 166803 (2009).

\bibitem{huang2015} H. Huang, Z. Liu, H. Zhang, W. Duan, and D. Vanderbilt,,
 Emergence of a Chern-insulating state from a semi-Dirac dispersion,
 Phys. Rev. B {\bf 92}, 161115(R) (2015.)

\bibitem{cai2015}Cai, T.,
  X. Li, F. Wang, S. Ju, J. Feng, and C.-D. Gong,
  Single-spin Dirac Fermion and Chern insulator based on
simple oxides, {\it Nano Lett.} {\bf 15}, 6434 (2015).



\bibitem{bacro3} H.-S. Jin, K.-H. Ahn, M.-C. Jung, and K.-W. Lee,
Strain and spin-orbit coupling induced orbital ordering in the 
Mott insulator BaCrO$_3$,
Phys. Rev. B {\bf 90}, 205124 (2014).

\bibitem{doennig2} D. Doennig, W. E. Pickett, and R. Pentcheva,
 Massive Symmetry Breaking in LaAlO$_3$/SrTiO$_3$(111) Quantum Wells: A
 Three-Orbital, Strongly Correlated Generalization of Graphene,
 Phys. Rev. Lett. {\bf 111}, 126804 (2013).

\bibitem{doennig} D. Doennig, W. E. Pickett, and R. Pentcheva,
 Confinement-driven transitions between topological and Mott phases
 in (LaNiO$_3$)$_N$/(LaAlO$_3$)$_M$(111) superlattices,
 Phys. Rev. B {\bf 89}, 121110(R) (2014).

\bibitem{fplo1} K. Koepernik and H. Eschrig,
 Full-potential nonorthogonal local-orbital minimum-basis band-structure scheme,
 Phys. Rev. B {\bf 59}, 1743 (1999).

\bibitem{wien2k} K. Schwarz and P. Blaha,
 Solid state calculations using WIEN2k,
 Comput. Mater. Sci. {\bf 28}, 259 (2003).


\bibitem{bfpo1} H. Kabbour, R. David, A. Pautrat, H.-J. Koo, M.-H. Whangbo, 
 G. Andr\'{e}, and O. Mentr\'{e},
 A genuine two-dimensional Ising ferromagnet with magnetically driven 
re-entrant transition,
 Angew. Chem. Int. Ed. {\bf 51}, 11745 (2012).

\bibitem{bfpo2} R. David, A. Pautrat, D. Filimonov, H. Kabbour, H. Vezin,
 M.-H. Whangbo, and O. Mentr\'{e},
 Across the structural re-entrant transition in \bfpo: influence 
 of the two-dimensional ferromagnetism,
 J. Am. Chem. Soc. {\bf 135}, 13023 (2013).

\bibitem{bfpo3} R. David, H. Kabbour, D. Filimonov, M. Huv\'{e},
 A. Pautrat, and O. Mentr\'{e},
 Reversible topochemical exsolution of iron in BaFe$^{2+}_2$(PO$_4$)$_2$,
 Angew. Chem. Int. Ed. {\bf 53}, 13365 (2014).

\bibitem{BFPO1} 
Y.-J. Song, K.-W. Lee, and W. E. Pickett,
 Large orbital moment and spin-orbit enabled Mott transition in the 
  Ising Fe honeycomb lattice of BaFe$_2$(PO$_4$)$_2$,  
Phys. Rev. B {\bf 92}, 125109 (2015).

\bibitem{BFPO2}
Y.-J. Song, K.-H. Ahn, W. E. Pickett, and K.-W. Lee,  
Tuning ferromagnetic BaFe$_2$(PO$_4$)$_2$ through a high Chern topological phase, 
Phys. Rev. B {\bf 94}, 125134 (2016).


\bibitem{xiao2011}
D. Xiao, W. Zhu, Y. Ran, N. Nagaosa, and S. Okamoto,
Interface engineering of quantum Hall effects in
digital transition metal oxide heterostructures,
\newblock Nat. Commun. {\bf 2}, 596 (2011).
\bibitem{cook} A. M. Cook and A. Paramekanti,
 Double Perovskite Heterostructures: Magnetism, Chern Bands, and Chern Insulators,
 Phys. Rev. Lett. {\bf 113}, 077203 (2014).

\bibitem{2LXOb}
D. Doennig, S. Baidya, W. E. Pickett, and R. Pentcheva,
Design of Chern and Mott insulators in buckled 3d-oxide honeycomb bilayers,  
Phys. Rev. B {\bf 93}, 165145 (2016).

\bibitem{2LXOa}H. Guo, S. Gangopadhyay, O. Koeksal, R. Pentcheva, and W. E. Pickett,  
Wide gap Chern Mott insulating phases achieved by design,  
npj Quantum Materials {\bf 2}, 4 (2017).

\end{thebibliography}
\end{document}